\documentclass[10pt]{article}
\usepackage{amsmath,amssymb,amsfonts}
\usepackage{units}
\usepackage{bbold, lmodern}%opcoes de fonte: mathptmx, helvet, eulervm, avant, fourier, newcent, mathpazo,euler,newcent,sansmath, lmodern
\usepackage{graphicx}% Include figure files
\usepackage{dcolumn}% Align table columns on decimal point
\usepackage{bm,bbm,mathtools,esvect}
\usepackage{slashed}
\usepackage{esvect}
\usepackage{fullpage}
\usepackage[utf8]{inputenc}
\usepackage{mathrsfs}
\title{ \bf COMPLEX LAGRANGIAN DYNAMICS}% \vspace{2cm}}
\author{\bf{SERGIO GIARDINO}\footnote{\tt sergio.giardino@ufrgs.br}\\ 
\\% \vspace{1cm} \\
\small \sf Departamento de Matem\'atica Pura e Aplicada \\
\small \sf Universidade Federal do Rio Grande do Sul -- UFRGS\\
\small \sf Caixa Postal 15080 -- 91501-970 -- Porto Alegre -- RS \\
\small \sf Brazil}
\begin{document}
\date{} % Remove a data
\maketitle
\newtheorem{theorem}{Theorem}[section] 
\newtheorem{remark}{Remark}[section] 
\newtheorem{lemma}{Lemma}[section] 
\newtheorem{proposition}{Proposition}[section] 
\newtheorem{corollary}{Corollary}[section] 
\newtheorem{definition}{Definition}[section]

%\vspace{1cm}

\begin{abstract}
	
In this article one introduces a formalism of classical mechanics where complex Lagrangian functions are admitted. The results include complex versions of the Lagrangian function, of the Euler-Lagrange equation, of the Hamilton principle, a geometric formulation, and the relation to a previous complex Hamiltonian formalism.  The framework is particularly suitable for non-stationary motion, and various pathways can be followed in future investigation.
\noindent 

\vspace{2mm}

\noindent keywords: formalism of classical mechanics%; quantum mechanics.

\vspace{1mm}

\noindent pacs numbers: 45.20.-d%; 03.65.-w.

\end{abstract}

\vspace{2cm}

\hrule
{\parskip - 0.3mm \footnotesize{\tableofcontents}}
\vspace{1cm}
\hrule

\pagebreak

%%%%%%%%%%%%%%%%%%%%%%%%%%%%%%%%%%%%%%%%%%%%%%%%%%%%%%%%%%%%%%%%%%%%%%%%%%
%%%%%%%%%%%%%%%%%%%%%%%%%%%%%%%%%%%%%%%%%%%%%%%%%%%%%%%%%%%%%%
\section{ Introduction\label{I}}
%%%%%%%%%%%%%%%%%%%%%%%%%%%%%%%%%%%%%%%%%%%%%%%%%%%%%%%%%%%%%%%
%%%%%%%%%%%%%%%%%%%%%%%%%%%%%%%%%%%%%%%%%%%%%%%%%%%%%%%%%%%%%%%%%%%%%%%%%%

The Hamiltonian formalism of classical mechanics \cite{Arnold:1989mma,Ratiu:2002isy,Abraham:2008fom} is stated through a symplectic structure that is similar to the proper structure of the complex numbers. However, the connections between classical mechanics and complex numbers may exceed the symplectic structure, and the  complex structure was examined in \cite{Strocchi:1966cca,Kramer:1975gta}  as a way to unify classical mechanics with quantum mechanics, an idea that was further explored by various authors, {\it e. g.} \cite{Heslot:1985qam,Jones:1992krw,Millard:1996qj,Prezhdo:1996gs,Gerdjikov:2002xd,Roncadelli:2007zz,Braasch:2022hjj}, although an accepted generalization is still lacking. More recently, a further attempt considers a complex phase space \cite{Giardino:2025jni}, where an interesting analogy between classical and quantum mechanics in the Hamiltonian formalism was established using the anti-symmetric scalar product, a formalism that permits considering particular dissipative models in a natural way, as well as the relation to Gromov's non-squeezing theorem \cite{deGosson:2009fop}.

These previous formulations use the Hamiltonian formulation as the intersection point between classical and quantum mechanics, but a Lagrangian complex approach is by far less explored than the Hamiltonian one. Complex Lagrangian functions are common in quantum field theory, and one can hypothesize whether the Lagrangian structure of classical mechanics also admits a complexification. Former attempts to this question involve K\"ahler manifolds and complex variables \cite{tekkoyun:2006org,cabar:2007clh}, imaginary time Lagrangian function \cite{Rami:2013iti}, and complex coordinates \cite{castillo:2023app} as well. In this article, one extends the prototype developed to the complex phase space \cite{Giardino:2025jni} to the Lagrangian language.

As will be clear in the sequel, the theory permits the description of dissipative mechanical systems in a simple way, a sound advantage of the formalism. There is an amount of research concerning dissipative mechanical systems \cite{Razavy:2005ecs,bersani:2021ldr}, as well as  speculations involving non-linear complex mechanics \cite{Bodurov:1998kk}, and the Herglotz theory \cite{Lazo:2017udy,deLeon:2020hnm,zhang:2020sym}. Comparing the complex Lagrangian theory introduced in this article to these previous ways to treat dissipative systems, as well as the more sophisticated contact geometric theory \cite{brea:2021mmm,Grabowski:2022geo}, Finsler geometry \cite{munteanu2004complex,El-Nabulsi:2021jnu}, and complex manifolds \cite{Bruzzo:2001cle,Lobos:2009prl} are of course fascinating directions for future research.
Nevertheless, one has to mention previous ways to complexify classical mechanics, including Hamiltonian formalism \cite{Bender:2006tz,Bender:2012yv,Bender:2009ab,Gorsky:2014lia,Marletto:2025egb,nigam2016quantum,el2012lagrangian}, Lagrangian formalism \cite{munteanu2004complex,el2012lagrangian} complex energy \cite{Bender:2006tz,Anderson:2012bf,Anderson:2011fn}, and fractional calculus \cite{El-Nabulsi:2009fqu,El-Nabulsi:2011efc}, but the theory presented here is theoretically simple, and admits a clear analogy to the elements of the real theory as well.

Summarily, in this paper one extended the previously developed complex Hamiltonian formalism of classical mechanics to a Lagrangian formulation \cite{Giardino:2025jni}. In principle, it seems an immediate consequence of the results contained in the previous article, but is more than that: the Lagrangian formalism is necessary in order to extend the complex formulation to quantum field theory, and naturally opens the possibility of developing second quantization methods. In other words, the results contained in this article are necessary in order to extend the range of the previous results. 

The article is organized as follows: the Lagrangian formalism is obtained after introducing the novel the dynamical equation (\ref{la04}), and consequently the generalized Euler-Lagrange equation (\ref{la06}), immediately used to study non-stationary mechanical systems (\ref{la16}), (\ref{la17}) and (\ref{la19}). These examples conform the previous Hamiltonian formulation that associates imaginary components to either dissipative or forced dynamical systems. The novel formalism has been submitted to consistency tests in sections considering equivalent Lagrangian functions, the Hamilton principle, and the geometric formulation. After assuring the confidence of the Lagrangian model and their internal consistency, the last section relates the novel formalism to the previously obtained Hamiltonian model \cite{Giardino:2025jni}. Therefore, all the sections present novel results, and one can start introducing the fundamental ideas in the following section.
%}
%%%%%%%%%%%%%%%%%%%%%%%%%%%%%%%%%%%%%%%%%%%%%%%%%%%%%%%%%%%%%%%%%%%%%%%%%%
%%%%%%%%%%%%%%%%%%%%%%%%%%%%%%%%%%%%%%%%%%%%%%%%%%%%%%%%%%%%%%%
\section{Complex Lagrangian function\label{La}}
%%%%%%%%%%%%%%%%%%%%%%%%%%%%%%%%%%%%%%%%%%%%%%%%%%%%%%%%%%%%%%%
%%%%%%%%%%%%%%%%%%%%%%%%%%%%%%%%%%%%%%%%%%%%%%%%%%%%%%%%%%%%%%%%%%%%%%%%%%
The usual Lagrangian dynamics comprise
\begin{equation}\label{la01}
 p=\frac{\partial L}{\partial \dot q},\qquad \dot p=\frac{\partial L}{\partial q},
\end{equation}
where $L=L(q,\,\dot q,\,t)$ is the real Lagrangian function, $q$ is the generalized position coordinate, $p$ is the generalized linear momentum coordinate, and the dot notation of course represents a time derivative. In order to obtain a complex formulation in the same token as the complex Hamiltonian formalism introduced in \cite{Giardino:2025jni}, one defines 
the complex variable $w$ and the derivative with respect to it
\begin{equation}\label{la03}
	w=\frac{1}{\sqrt{2}}\big(\dot q+ \omega_0 q\, i\big),\qquad\mbox{and}\qquad \frac{\partial}{\partial w}=\frac{1}{\sqrt{2}}\left(\frac{\partial}{\partial\dot q}-\frac{i}{\omega_0}\frac{\partial}{\partial q}\right),
\end{equation}
where $\omega_0$ is a dimensional constant. After introducing the complex variable $u$, so that
\begin{equation}
	u=\frac{1}{\sqrt{2}}\big(\dot p+ \omega_0 p\, i\big),
\end{equation}
the usual Lagrangian equations (\ref{la01}) are immediately obtained as the real components of the complex equation
\begin{equation}\label{la02}
 u=i\omega_0\frac{\partial L}{\partial w}.
\end{equation}
One notices that the above formulation does not involve holomorphic functions, and thus $L$ is not a complex variable function usually considered in complex analysis. A possible relation between the complex variables can be established in case of point particles of mass $m$, such that
\begin{equation}
u=m\dot w,
\end{equation}
and therefore,
\begin{equation}
	\dot w=i\frac{\omega_0}{m}\frac{\partial L}{\partial w},
\end{equation}
thus establishing a perfect parallel to the complex Hamiltonian  dynamics formulation \cite{Giardino:2025jni}. In accordance with this complex Hamiltonian formalism, one defines the complex Lagrangian function $\mathfrak L$, such as
\begin{equation}\label{la15}
  \mathfrak L=L+Mi
\end{equation}
where $L$ and $M$ are of course real functions. The complex expression of the Lagrangian dynamics in (\ref{la02}) is also defined to be
\begin{equation}\label{la04}
u=i\omega_0\frac{\partial\mathfrak L }{\partial w}
\end{equation}
leading to the dynamical equations
\begin{equation}\label{la05}
 p=\frac{\partial L}{\partial \dot q}+\frac{1}{\omega_0}\frac{\partial M}{\partial q},\qquad \dot p=\frac{\partial L}{\partial  q}-\omega_0\frac{\partial M}{\partial \dot q}.
\end{equation}
As expected, the limit  $M\to 0$ immediately recovers the usual Lagrangian dynamics (\ref{la01}). At this moment, one observes (\ref{la15}) as a seemingly arbitrary mathematical generalization, where the function $M$ has been imposed to the Lagrangian function without physical motivation. However, these motivations will be clear after considering the consequences of this complex generalization. First of all, one observes that the dynamical equations (\ref{la05}) immediately give
\begin{equation}\label{la06}
\frac{\partial L}{\partial q}- \frac{d}{dt}\frac{\partial L}{\partial \dot q}=\omega_0\frac{\partial M}{\partial\dot q}+\frac{1}{\omega_0} \frac{d}{dt}\frac{\partial M}{\partial q},
\end{equation}
and (\ref{la06}) recovers the usual Euler-Lagrange equation in the limit $M\to 0$, as expected. A nonzero right hand side of (\ref{la06}) hypothesizes  non-conservative mechanical processes described within this formalism. In order to verify the physical character of the imaginary component $M$ of the Lagrangian function, one considers several examples.

%%%%%%%%%%%%%%%%%%%%%%%%%%%%%%%%%%%%%%%%%%
\section{Complex Lagrangian systems}
%%%%%%%%%%%%%%%%%%%%%%%%%%%%%%%%%%%%%%%%%%

In this section, one considers several conceptually interesting possibilities.

\paragraph{\underline{Inverted harmonic oscillator}} In order to evaluate the physical consequences of the complex Lagrangian functions, one considers a pure imaginary Lagrangian function, obtained as an analog of  an harmonic oscillator of mass $m$ and elastic constant $k$, where (\ref{la15}) reads
\begin{equation}\label{la16}
\mathfrak L	=\frac{i}{2}\Big(m\dot q^2-kq^2\Big),
\end{equation}
and the equations of motion obtained from (\ref{la05}) are immediately obtained 
\begin{equation}
p=-\frac{k}{\omega_0}q,\qquad\qquad \dot p=-\omega_0 m\dot q,
\end{equation}
and whose solution is
\begin{equation}\label{la08}
	q=q_0\exp\big[\omega_0 t\big],\qquad \mbox{where}\qquad \omega_0^2=\frac{k}{m},
\end{equation}
with $q_0$ as the integration constant.
Consequently, showing that (\ref{la16}) corresponds either to a dissipative or to a forced non-oscillating physical process, depending on the signal of $\omega_0$. One can get a further comprehension from (\ref{la08}), so that
\begin{equation}\label{la09}
q=\omega_0^2q,
\end{equation}
is easily recognized as the equation of the well-known inverted harmonic oscillator. Therefore, the imaginary component of the Lagrangian function may generate equations of motion of a dissipative or forced physical process. This is an easy way to obtain equations of motion for such kind of process, and whose current theory may actually be rather involved \cite{Razavy:2005ecs,bersani:2021ldr}.

\paragraph{\underline{Harmonic oscillator}} The previous example demonstrates the completely different character that the real  and imaginary components of a complex Lagrangian function may assume. Let us then consider
\begin{equation}\label{la17}
	\mathfrak L	=i\alpha_0 q \dot q,
\end{equation}
where $\alpha_0$ is a dimensional constant. The equations of motion generated from (\ref{la05}) conforms to
\begin{equation}
\dot p=-\omega_0\alpha_0 q,\qquad\mbox{and}\qquad p=\frac{\alpha_0}{\omega_0}\dot q.
\end{equation}
These equations permit to get
\begin{equation}\label{la18}
\ddot q=-\omega_0^2q\qquad \mbox{and}\qquad \alpha_0=m\omega_0,
\end{equation}
the well-known equation of an harmonic oscillator, indicating a curious symmetry between imaginary and real component of a complex Lagrangian function, because multiplying (\ref{la16}) by $i$, one generate a real Lagrangian function whose equation of motion is (\ref{la18}), and a similar relation exists between (\ref{la17}) and (\ref{la09}). The possible implications of this symmetry are interesting directions for future research. However, based on these two simple examples, one can propose a further example.
 
%%%%%%%%%%%%%%%%%%%%%%%%%%%%%%%%%%%%%%%%%%%%%%%%%%%%%%%%%%
\paragraph{\underline{Non-stationary oscillator}}
%%%%%%%%%%%%%%%%%%%%%%%%%%%%%%%%%%%%%%%%%%%%%%%%%%%%%%%%%%
Let us then consider the Lagrangian function
\begin{equation}\label{la19}
	\mathfrak L	=\frac{i}{2}\Big(m\dot q^2-kq^2\Big)+\frac{i}{2}\lambda_0  \dot q^2,
\end{equation}
where $\lambda_0$ is a dimensional constant. The equations of motion thus read
\begin{equation}
	p=m\dot q,\qquad\qquad \ddot q+\omega_0^2 q+\omega_1\dot q=0,
\end{equation}
where one identifies $\omega_1=\lambda_0\omega_0/m$. The above equation describes a non-stationary oscillation. To the best of our knowledge, this is the simplest way to generate a non-stationary process from a Lagrangian function, and the directions for future investigation are exciting.

%%%%%%%%%%%%%%%%%%%%%%%%%%%%%%%%%%%%%%%%%%%%%%%%%%%%%%%%%%
\section{Equivalent Lagrangian functions}
%%%%%%%%%%%%%%%%%%%%%%%%%%%%%%%%%%%%%%%%%%%%%%%%%%%%%%%%%%
It is a well-known fact of classical mechanics that a particular Lagrangian function uniquely determines the equations of motion, but the reciprocal are no true, and the equations of motion do not determine the Lagrangian function uniquely. Conversely, if $L_1$ and $L_2$ determine a particular set of equations of motion, it holds that the difference between these functions is so that
\begin{equation}\label{la20}
	L_2-L_1=\frac{d\Lambda}{dt},
\end{equation}
where $\Lambda$ is a real function of the time, of the coordinates of the problem, and of their first derivatives. In order to ascertain whether a similar result holds for the complex formulation determined in previous sections, let us consider two complex Lagrangian functions, determined by the real functions $L_1,\,M_1,\,L_2$ and $M_2$. Expanding the total derivative of  (\ref{la06}) on both of the cases permits one to obtain
\begin{equation}\label{la07}
 \frac{\partial F}{\partial \dot q}\ddot q +\frac{\partial F}{\partial q}\dot q+\frac{\partial F}{\partial t}-\frac{\partial \Delta L}{\partial q}+\omega_0\frac{\partial \Delta M}{\partial \dot q}=0
\end{equation}
where
\begin{equation}
F=\frac{\partial\Delta L}{\partial \dot q}+\frac{1}{\omega_0}\frac{\partial\Delta M}{\partial q},
\end{equation}
and accordingly
\begin{equation}  
\qquad \Delta L=L_2-L_1,\qquad\mbox{and}\qquad \Delta M=M_2-M_1.
\end{equation}
The original real Lagrangian function condition (\ref{la20}) is obtained if $\Delta M=0$, as expected. On the other hand, $\Delta L=0$  forces $\Delta M=0$ as well, indicating that (\ref{la20}) seems reasonable even in the complex case.

Nevertheless, if expects a more general situation, where both $\Delta L$ and $\Delta M$ are not identically zero, one observes  the first term of (\ref{la07}) forcefully to be zero, because the second derivative $\ddot q$ does not appear in the remaining terms, and evidently $F$ does not depend on $\ddot q$. Considering this condition, the partial derivative relative to $\dot q$ of the whole (\ref{la07}) equation renders
\begin{equation}
\frac{\partial}{\partial\dot q}\left(\frac{\partial \Delta L}{\partial q}-\omega_0\frac{\partial \Delta M}{\partial \dot q}\right)=0,
\end{equation}
and this condition forces the second term of (\ref{la07}) also to be identically zero, so that
\begin{equation}
	F=F(t).
\end{equation}
Therefore, (\ref{la07}) goes to
\begin{equation}
\frac{\partial}{\partial \dot q}\left(\frac{\partial\Delta L}{\partial t}+\omega_0\Delta M\right)-
\frac{\partial}{\partial  q}\left(\Delta L-\frac{1}{\omega_0}\frac{\partial\Delta M}{\partial t}\right)=0,
\end{equation}
whose integrability condition reads
\begin{equation}
\frac{\partial\Phi}{\partial q}=\frac{\partial\Delta L}{\partial t}+\omega_0\Delta M\qquad \mbox{and}\qquad 
\frac{\partial\Phi}{\partial  \dot q}=\Delta L-\frac{1}{\omega_0}\frac{\partial\Delta M}{\partial t},
\end{equation}
where $\Phi=\Phi(q,\,\dot q,\,t)$. After some algebra, one finally obtains
\begin{equation}\label{la32}
	\left(\frac{\partial^2}{\partial t^2}+\omega_0^2\right)F=\left(\frac{\partial^2}{\partial q^2}+\omega_0^2\frac{\partial^2}{\partial \dot q^2}\right)\Phi.
\end{equation}
In the simplest case, where $\Phi=0$, one obtains $F$ to be described in terms of an harmonically oscillating  function. This is a very curious result, and seems to be an interesting direction for future investigation.

%%%%%%%%%%%%%%%%%%%%%%%%%%%%%%%%%%%%%%%%%%%%%
\section{Hamilton's principle}
%%%%%%%%%%%%%%%%%%%%%%%%%%%%%%%%%%%%%%%%%%%%%

The physical trajectory $q$ in configuration space of a mechanical system minimizes the action integral
\begin{equation}\label{la21}
	S=\int_{t_0}^{t_1} L \,dt,
\end{equation}
where $L=L(q,\,\dot q,\,t)$ is the usual Lagrangian function. This statement is the well known Hamilton's principle of classical mechanics, and one has to investigate whether a similar principle holds to the complex Lagrangian function. Therefore, one replaces the real action (\ref{la21}) by
\begin{equation}\label{la34}
	\mathcal S=\int_{t_0}^{t_1} \mathfrak L \,dt,
\end{equation}
where $\mathfrak L$ is the complex Lagrangian function (\ref{la15}). The infinitesimal variation of the complex action consequently reads
\begin{equation}\label{la13}
 \delta\mathcal S=\int\delta \mathfrak L\, dt
\end{equation}
where
\begin{equation}\label{la14}
 \delta\mathfrak L=\frac{\partial \mathfrak L}{\partial w} \delta w,
\end{equation}
and the complex variable $w$ is defined according to (\ref{la03}). Therefore,
\begin{equation}
 \delta\mathfrak L=\left(\frac{\partial L}{\partial q}-\omega_0\frac{\partial M}{\partial \dot q}\right)\delta q+
 \left(\frac{\partial L}{\partial \dot q}+\frac{1}{\omega_0}\frac{\partial M}{\partial q}\right)\delta \dot q+
 i\left[\left(\omega_0\frac{\partial L}{\partial \dot q}+\frac{\partial M}{\partial q}\right)\delta  q-
 \left(\frac{1}{\omega_0}\frac{\partial L}{\partial q}-\frac{\partial M}{\partial \dot q}\right)\delta \dot q
  \right].
\end{equation}
After integration by parts, the equation of motion (\ref{la06}) is obtained in the condition that
\begin{equation}\label{la22}
 \mathfrak{Re}[\delta\mathcal S]=0,
\end{equation}
remembering that $\delta q=0$ in the limit points of the integration. The reality condition (\ref{la22}) can be further understood in terms of the inner product defined in \cite{Giardino:2025jni} as
\begin{equation}\label{la23}
(z,\,w)=\mathfrak{Re}[z\,\overline w],
\end{equation}
and one can thus write
\begin{equation}\delta\mathfrak L=\left(\frac{\partial \mathfrak L}{\partial w},\, \delta \overline w\right).
\end{equation}
The interpretation of the variation of the action in terms of an inner product is already known even in terms of the real theory. The concordance in this point further indicates the consistence of the theory. Also a generalization of the Noether theorem is admitted, considering an arbitrary transformation to require
\begin{equation}
\frac{d}{dt}\left[\left(\frac{\partial L}{\partial \dot q}+\frac{1}{\omega_0}\frac{\partial M}{\partial q}\right)\delta q\right]=0,
\end{equation}
and consequently
\begin{equation}\label{la33}
\Gamma= \left(\frac{\partial L}{\partial \dot q}+\frac{1}{\omega_0}\frac{\partial M}{\partial q}\right)\delta q
\end{equation}
is a constant of the motion. The interpretation, however, is not so simple as in the case of the real theory, because the conservative momentum in the real theory is coherent with a Lagrangian function independent of the coordinate, and this is not observed in (\ref{la06}) after imposing $\partial L/\partial q=\partial M/\partial q=0$. The precise physical meaning of the Noether theorem in this complex formalism is a fascinating direction for future research. 

On the other hand, one does not expect the characteristic symmetry of the real Lagrangian formalism to be found in the complex Lagrangian formulation because the inclusion of non-stationary systems in the complex formalism possibly hinders certain symmetry transformations exclusive for stationary systems.   One immediately observes that an $i$ factor relates the inverted oscillator Lagrangian function (\ref{la16}) to the usual harmonic oscillator real Lagrangian function. Therefore, a linear combination of complex Lagrangian functions requires real coefficients, and equations of motion obtained from the complex action (\ref{la34}) are invariant  by a multiplication of a real factor, but there equations of motion are not invariant after the multiplication by a complex factor. The necessity of real coefficients has already been observed in the Hamiltonian formulation of the phase space \cite{Giardino:2025jni}, and this idea also appears in the real Hilbert space quantum mechanics \cite{Giardino:2018rhs}. Exploring the coherence between the classical and quantum formulations is an exciting direction of future research.

Eventually, equivalent equations of motion could be represented by different Lagrangian functions, which may possibly be related by symmetry transformations,  and the multiplication by a constant complex or real factor is only the simplest possibility. One finally remembers  the role of symmetry operations as non-exhausted  subject even in the real classical mechanics, as one can observe from some recent research work \cite{Aoki:2022ugd,Azuaje:2023zti,castillo:2014osy,Cariglia:2014ysa}, and the novel possibilities introduced by the complex formalism can open a novel way for future investigation of this subject 
%}

%%%%%%%%%%%%%%%%%%%%%%%%%%%%%%%%%%%%%%%%%%%%%%%%%%%%%%%%%%
\section{Geometric formulation}
%%%%%%%%%%%%%%%%%%%%%%%%%%%%%%%%%%%%%%%%%%%%%%%%%%%%%%%%%%
In this section, one can ascertain the complex formalism to be compatible with a geometric formulation. In order to accomplish that, one remembers the Lie derivative as a differential operator  defined in terms of a vector field $\bm X$, so that
\begin{equation}\label{g01}
 \widehat L_{\bm X}=X_1^a\frac{\partial}{\partial q^a}+X_1^a\frac{\partial}{\partial \dot q^a},
\end{equation}
where
\begin{equation}\label{g02}
 \bm X=\big(X_1^a,\,X_2^a\big).
\end{equation}
A Lagrangian vector field reads
\begin{equation}\label{g03}
 \bm\Delta=\big(\dot q^a,\,\ddot q^{\,a}\big),
\end{equation}
and the Lie derivative in this case can be identified with a total derivative. One may stipulate the Lagrangian one-form to be
\begin{equation}\label{g04}
 \Theta_L=\left(\frac{\partial L}{\partial\dot q^a}+\frac{1}{\omega_0}\frac{\partial M}{\partial q^a}\right)dq^a,
\end{equation}
applying the Lie derivative over the Lagrangian one-form, and using the Euler-Lagrange equation (\ref{la06}), one obtains
\begin{equation}\label{g05}
\widehat L_{\bm\Delta}\Theta_L=\left(\frac{\partial L}{\partial q^a}-\omega_0\frac{\partial M}{\partial\dot q^a}\right)dq^a +
\left(\frac{\partial L}{\partial\dot q^a}+\frac{1}{\omega_0}\frac{\partial M}{\partial q^a}\right)d\dot q^a.
\end{equation}
As expected, $M=0$ recovers the usual real result, where the right hand side of (\ref{g05}) is exactly the differential one-form $dL$ within this real limit. However, in the $M\neq 0$ case the right hand side is not something like $d\mathfrak L$, because both of the sides have to be real. The correct equality is obtained using the inner product (\ref{la23}), and thus
\begin{equation}
\widehat L_{\bm\Delta}\Theta_L=2\left(\frac{\partial\mathfrak L}{\partial w},\,dw\right).
\end{equation}
This result provides an interesting insight into complex geometry that is compatible to \cite{Giardino:2025jni}, and further developments are stimulating routes for investigation.

%%%%%%%%%%%%%%%%%%%%%%%%%%%%%%%%%%%%%%%%%%
\section{Relation to Hamiltonian dynamics}
%%%%%%%%%%%%%%%%%%%%%%%%%%%%%%%%%%%%%%%%%%

This result relates to a previous Hamiltonian description \cite{Giardino:2025jni},
\begin{equation}\label{la24}
 \dot q=\frac{\partial H}{\partial p}-\varkappa_0\frac{\partial K}{\partial q},\qquad \dot p=-\frac{\partial H}{\partial q}-\frac{1}{\varkappa_0}\frac{\partial K}{\partial p}.
\end{equation}
where $\varkappa_0$ is a dimensional constant to be chosen according to the specific physical situation. 
In order to relate the results, one change the coordinates of $L$ and $M$ such as
\begin{equation}\label{la25}
 \mathsf L=L\big(q,\,\dot q(q,\,p,\,t),\,t\big),\qquad \mathsf M=M\big(q,\,\dot q(q,\,p,\,t),\,t\big)
\end{equation}
Therefore, it holds that
\begin{equation}\label{la26}
 \frac{\partial\mathsf L}{\partial q}=\frac{\partial L}{\partial q}+\frac{\partial L}{\partial \dot q}\frac{\partial \dot q}{\partial q},\qquad\mbox{and}\qquad
  \frac{\partial\mathsf L}{\partial p}=\frac{\partial\dot q}{\partial p}\frac{\partial L}{\partial\dot q},
\end{equation}
as well as to $\mathsf M$. Eliminating the derivatives of $L$ using (\ref{la05}) leads to
\begin{equation}\label{la27}
\dot p=-\frac{\partial H}{\partial q}+\frac{1}{\omega_0}\frac{\partial M}{\partial\dot q}-\omega_0\frac{\partial M}{\partial q}\frac{\partial \dot q}{\partial q}
\end{equation}
where the usual definition of the Hamiltonian function holds, namely
\begin{equation}\label{la28}
 H=p\dot q-\mathsf L.
\end{equation}
Repeating to $M$ the process done to $L$ to change the variables using (\ref{la26}), one obtains
\begin{equation}\label{la29}
 \dot p=-\frac{\partial H}{\partial q}+\frac{1}{\omega_0}\frac{\partial \dot q}{\partial q}\frac{\partial \mathsf M}{\partial q}-
 \frac{\omega_0+\frac{1}{\omega_0}\left(\frac{\partial \dot q}{\partial q}\right)^2}{\frac{\partial \dot q}{\partial p}}\frac{\partial \mathsf M}{\partial p}.
\end{equation}
A comparison to (\ref{la24}) implies
\begin{equation}
	\frac{\partial K}{\partial p}=\varkappa_0\left(-\frac{1}{\omega_0}\frac{\partial \dot q}{\partial q}\frac{\partial \mathsf M}{\partial q}+
	\frac{\omega_0+\frac{1}{\omega_0}\left(\frac{\partial \dot q}{\partial q}\right)^2}{\frac{\partial \dot q}{\partial p}}\frac{\partial \mathsf M}{\partial p}\right).
\end{equation}
A similar procedure has to be done to obtain the derivative of $K$ relative to $q$. A similar procedure as that used to obtain (\ref{la27}) permits to obtain
\begin{equation}\label{la30}
 \dot q=\frac{\partial H}{\partial p}-\frac{1}{\omega_0}\frac{\partial\dot q}{\partial p}\frac{\partial M}{\partial q}.
\end{equation}
Changing the variables to eliminate the dependence on $\dot q$ renders
\begin{equation}
	\dot q=\frac{\partial H}{\partial p}+\frac{1}{\omega_0}\left(\frac{\partial \dot q}{\partial q}\frac{\partial\mathsf M}{\partial p}-\frac{\partial \dot q}{\partial p}\frac{\partial\mathsf M}{\partial q}\right)
\end{equation}
as well as
\begin{equation}\label{la31}
 \frac{\partial K}{\partial q}=\frac{1}{\varkappa_0\omega_0}\left(\frac{\partial \dot q}{\partial p}\frac{\partial\mathsf M}{\partial q}-\frac{\partial \dot q}{\partial q}\frac{\partial\mathsf M}{\partial p}\right),
\end{equation}
and hence the comparison to the Hamiltonian formalism is fully established. One can imagine a simpler way to establish such a correspondence, maybe using a complex formalism, but this is of course another fascinating track  for subsequent research.

%%%%%%%%%%%%%%%%%%%%%%%%%%%%%%%%%%%%%%%%%%%%%%%%%%%%%%%%%%%%%%%%%%%%%%%%%%
%%%%%%%%%%%%%%%%%%%%%%%%%%%%%%%%%%%%%%%%%%%%%%%%%%%%%%%%%%%%%%%
\section{ Conclusion\label{C}}
%%%%%%%%%%%%%%%%%%%%%%%%%%%%%%%%%%%%%%%%%%%%%%%%%%%%%%%%%%%%%%%
%%%%%%%%%%%%%%%%%%%%%%%%%%%%%%%%%%%%%%%%%%%%%%%%%%%%%%%%%%%%%%%%%%%%%%%%%%

In this article, a complex expression of the Lagrangian formulation of classical mechanics has been considered, and the equations of motion, the Hamilton principle, and the geometric formulation have been consistently built. Several simple examples have accordingly been considered, and it has been observed that the imaginary part of the Lagrangian can be related to non-stationary processes, a interpretation that is also consistent to the complex Hamiltonian formulation. Moreover, the complex Lagrangian formulation has been proven to be compatible to the complex Hamiltonian formulation, stressing the compatibility of both of the formalisms.

One could question about the concrete novelties introduced within this complex version of classical dynamics when compared to previous complexifications mentioned in the introduction of the paper. Initially, one could mention the mathematical simplicity depicted in Section 2, consequently facilitating the physical interpretation. Moreover, one can mention the similarity to the real theory. The Lagrangian and Hamiltonian formalism are clearly related in the proposed theory, also depicting as a positive character. The real classical mechanics is also generalized because of the admission of non-stationary and non-conservative processes in a natural way, in conformity to most of the different complex formulations. Finally, the theory proposed in this article may be used as a prototype to unify classical and quantum mechanics because of the common mathematical structure behind the real inner product used to prove Hamilton's principle, and that is also found in real Hilbert space quantum mechanics \cite{Giardino:2018rhs}, and this constitutes a fascinating direction of future research.

After establishing the foundations of both of the expressions, a broad way of possibilities can be explored, particularly involving non-stationary, dissipative and forced physical processes. The comprehension of these physical systems required Lagrangian and Hamiltonian formulations in order to be studied in the same basis as non-dissipative systems, and the results presented in this article seemingly  provide a way to establish such a fundamental basis. One expects that the application of this formalism may contribute to a deeper understanding of usual classical mechanics, and a mathematical framework to establish a dialogue to quantum mechanics, where the complex numbers are ubiquitous, and where novel descriptions of non-stationary solutions \cite{Giardino:2024tvp} have been recently considered in the real Hilbert state formalism \cite{Giardino:2018rhs}.

%%%%%%%%%%%%%%% Considerar transformacoes ativas e passicas [saletan 3.2.1], o teorema de noether [3.2.2] e o oscilador amortecido [3.3.2], para mostar a formulacao complexa como generalizacao deste caso. Tambem o [3.1.2], envolvendo vinculos, poderia ser considerado.

\begin{footnotesize}
\paragraph{Funding} The author gratefully thanks for the financial support by Fapergs under the grant 23/2551-0000935-8 within Edital 14/2022.

\paragraph{Data availability statement}The author declares that data sharing is not applicable to this article as no data sets were generated or analyzed during the current study.

\paragraph{Declaration of interest statement} The author declares that he has no known competing financial interests or personal relationships that could have appeared to influence the work reported in this paper.
\end{footnotesize}
%%%%%%%%%%%%%%%%%%%%%%
%
%
%  BIBLIOGRAPHY
%
%
\begin{footnotesize}

%	\bibliographystyle{unsrt} 
%	\bibliography{bib_MCH_2}
\end{footnotesize}
\end{document}